\documentclass[12pt]{iopart}

\def  \be   {\begin{equation}}
\def  \ee   {\end{equation}}
\def  \beq  {\begin{eqnarray}}
\def  \eeq  {\end{eqnarray}}
\def  \etal   {et~al.~}

\begin{document}

\title[Dynamo action in Gaussian-correlated helical turbulence]
{Amplification of magnetic fields by dynamo action in Gaussian-correlated helical turbulence}

\author{Leonid Malyshkin$^1$ and Stanislav Boldyrev$^2$}

\address{$^1$ Department of Astronomy and Astrophysics,
University of Chicago, 5640 S. Ellis Ave., Chicago, IL 60637}
%\ead{leonmal@uchicago.edu}
\address{$^2$ Department of Physics, University of Wisconsin-Madison, 
1150 University Ave., Madison, WI 53706}

\ead{leonmal@uchicago.edu}
%\ead{boldyrev@wisc.edu}

%------------------------------------------------------------------------------------------

\begin{abstract}
We investigate the growth and structure of magnetic fields amplified by kinematic dynamo action 
in turbulence with non-zero kinetic helicity. We assume a simple Gaussian velocity correlation 
tensor, which allows us to consider very large magnetic Reynolds numbers, up to $10^{12}$. 
We use the kinematic Kazantsev-Kraichnan model of dynamo and find a complete numerical 
solution for the correlation functions of growing magnetic fields.
\end{abstract}
%\pacs{}

%\submitto{\PS}

%\maketitle

%\normalsize

%------------------------------------------------------------------------------------------

\section{Introduction}

The prevailing theory for the origin of cosmic magnetic fields in the Universe 
is magnetic dynamo action, which is random stretching of magnetic field lines 
by motion of highly conducting plasmas or fluids in which these lines are 
frozen (e.g., Brandenburg and Subramanian 2005; Kulsrud 2005; Kulsrud and Zweibel 2008; 
Lynden-Bell 1994; Parker 1979; Vainshtein and Zeldovich 1972; Zweibel and Heiles 1997).
Magnetic fields in astrophysical systems are often observed to be correlated at scales 
significantly larger than the correlation scales of plasma motions. 
Such large-scale magnetic fields can be generated by 
dynamo action if the velocity field ${\bf v}({\bf x, t})$ possesses nonzero kinetic 
helicity $H=\int {\bf v}\cdot ({\nabla \times {\bf v}})\, d^3 x\neq 0$ 
(Moffatt 1978; Steenbeck \etal 1966).

In this work we explore the growth and structure of magnetic fields generated by  
dynamo action driven by a velocity field with non-zero kinetic helicity, in the 
framework of the kinematic Kazantsev-Kraichnan model. For simplicity, we assume a Gaussian 
correlation tensor of velocity fluctuations, which means that all velocity fluctuations 
are on a single-scale, that is, the turbulence consists of similar-size eddies. This assumption, 
although not commonly satisfied in astrophysical systems, allows us to consider very 
large magnetic Reynolds numbers, up to $10^{12}$. More general settings can be addressed    
by the same method, and will be investigated elsewhere. Fast dynamo action driven by 
Gaussian-correlated velocity field with non-zero helicity was first studied numerically 
in the framework of the Kazantsev-Kraichnan model by Kim and Hughes (1997). Their 
work was limited to a study of magnetic field generation at very small scales, much 
smaller than the velocity correlation scale. In particular, they focused on the 
dependence of the fastest growth rate of the field on the value of the magnetic Reynolds 
number. Our work is complementary to their work. We consider magnetic field 
generation and find field growth rates at {\it all} scales. Thus, we present a complete 
numerical solution for helical dynamo action. In addition, a different numerical method 
that we use to solve the helical dynamo equations, allows us to consider magnetic 
Reynolds numbers much larger than whose investigated by Kim and Hughes (1997).\footnote{
Kim and Hughes (1997) considered values of magnetic Reynolds number ${\rm Rm}$ up to 
$10^5$. We use ${\rm Rm}$ up to $10^{12}$, restricted only by a limited numerical
precision $\sim 10^{-16}$ of floating-point numbers of double type in our computer 
code.}
In the next section we describe the Kazantsev-Kraichnan model. In the final section, 
we present our results and give our conclusions.

\section{Kazantsev-Kraichnan model of magnetic dynamo action}

We use the Kazantsev-Kraichnan model of kinematic dynamo action in homogeneous and isotropic 
turbulence (Kazantsev 1968; Kraichnan 1968). In this model, the ensemble statistics of 
velocity fluctuations is assumed to be Gaussian, with zero mean, $\langle{\bf v}\rangle=0$, 
and the covariance tensor
\beq 
\langle {v^i}({\bf x},t){v^j}({\bf x}',t') \rangle \!=\!
\kappa^{ij}({\bf x}-{\bf x}')\delta(t-t'),
\label{V_V_TENSOR}
\eeq
where $\kappa^{ij}$ is an isotropic tensor of turbulent diffusivity,
\beq
\kappa^{ij}({\bf x})\!=\!\kappa_N 
\left(\delta^{ij}-\frac{x^ix^j}{x^2}\right)+
\kappa_L \frac{x^ix^j}{x^2}+g\epsilon^{ijk}x^{k}.  
\label{KAPPA}
\eeq
Here $\langle \rangle$ denotes ensemble average, 
$\epsilon^{ijk}$ is the unit anti-symmetric pseudo-tensor and
summation over repeated indices is assumed. 
The first two terms at the right-hand side of (\ref{KAPPA})
represent the mirror-symmetric, non-helical part, while function $g(x)$
describes the helical part of the velocity fluctuations. 
For an incompressible velocity field (the only case we consider here), 
we have $\kappa_N(x)=\kappa_L(x)+x\kappa'_L(x)/2$, where the prime 
denotes  derivative with respect to~$x=|{\bf x}|$. Therefore, 
to describe the velocity field, we specify only two independent functions, 
$\kappa_L(x)$ and $g(x)$. 

The magnetic field correlator can similarly be expressed as
\begin{eqnarray}
\langle B^i({\bf x}, t)B^j(0,t)\rangle = 
M_N\left(\delta^{ij}-\frac{x^ix^j}{x^2}\right) 
+ M_L\frac{x^ix^j}{x^2}+K\epsilon^{ijk}x^k,
\label{B_B_TENSOR}
\end{eqnarray} 
where the field solenoidality constraint ${\rm div}\,{\bf B}=0$ implies 
$M_N(x,t)=M_L(x,t)+(x/2)M'_L(x,t)$. To fully describe the magnetic correlator, 
we therefore need to find only two functions, $M_L(x, t)$ 
and $K(x, t)$, corresponding to magnetic energy and magnetic helicity. 
The Fourier transformed version of equation~(\ref{B_B_TENSOR}) is 
\beq
\langle B^i({\bf k},t)B^{*j}({\bf k},t)\rangle =
F_B(k,t)\left(\delta^{ij}-\frac{k^ik^j}{k^2}\right) 
- i\frac{H_B(k,t)}{2k^2}\epsilon^{ijl}k^l,
\label{MAGNETIC_SPECTRA}
\eeq
where $F_B(k,t)$ is the magnetic energy spectral function, 
$\langle|{\bf B}({\bf k},t)|^2\rangle=2F_B(k,t)$, and $H_B(k,t)$
is the spectral function of the electric current helicity,
$\langle {B^i}^*({\bf k},t)\: i\epsilon^{ijl}k^jB^l({\bf k},t)\rangle=H_B(k,t)$.
Functions $F_B(k,t)$ and $H_B(k,t)$ can be obtained from functions $M_L(x,t)$ 
and $K(x, t)$, and vice verse, by using the three-dimensional Fourier 
transforms (Monin and Yaglom 1971).
The problem is then to find the correlation function~(\ref{B_B_TENSOR}) of the 
magnetic field, or, alternatively, its Fourier version~(\ref{MAGNETIC_SPECTRA}). 

The growing magnetic field ${\bf B}({\bf x, t})$, amplified by dynamo action, 
satisfies the induction equation 
\begin{eqnarray}
\partial_t {\bf B}=\nabla \times ({\bf v}\times {\bf B})+\eta \nabla^2 {\bf B},
\end{eqnarray}
where $\eta$ is the magnetic diffusivity. 
Suppose that the velocity field~(\ref{V_V_TENSOR})--(\ref{KAPPA}) is given, 
i.e.~kinetic energy~$\kappa_L(x)$ and kinetic helicity~$g(x)$ are given. 
It turns out that in this case, to find the properties of the growing magnetic 
field, one needs to solve two coupled linear homogeneous partial differential 
equations for functions $M_L(x,t)$ and $K(x, t)$ related to magnetic energy and 
magnetic helicity. These equations were first derived by Vainshtein and Kichatinov (1986) 
in the framework of the Kazantsev-Kraichnan model with non-zero kinetic helicity.
Recently, Boldyrev \etal (2005) established that the system of Vainshtein-Kichatinov 
equations possesses a self-adjoint structure, which is somewhat similar to a two-component 
quantum mechanical ``spinor'' form with imaginary time. We solve these two self-adjoint 
linear homogeneous partial differential equations numerically by the fourth-order 
Runge-Kutta integration method and by matching the numerical solution to the 
analytical asymptotic solutions at $x\to0$ and $x\to\infty$, for details see 
Malyshkin and Boldyrev (2007). 
A general solution for the growing magnetic field is a linear superposition of 
exponentially growing eigenmode solutions. Each eigenmode solution grows in time as 
$\exp(\lambda t)$, where $\lambda$ is the eigenmode growth rate. The self-adjoint 
structure of Boldyrev \etal (2005) equations guarantees that all growth rates are 
real. We find that, in analogy with quantum mechanics, there are two types of field 
eigenmodes: bound (spatially localized) and unbound (spatially non-localized). 
First, for growth rates $\lambda>\lambda_0\equiv g^2(0)/[\kappa_L(0)+2\eta]$ the 
eigenfunctions are bound and correspond to ``particles'' trapped by the potential 
provided by velocity fluctuations. The bound eigenmodes have discrete 
growth rates, i.e., $\lambda=\lambda_n>\lambda_0$, $n=1,2,3...$. 
As $x\to\infty$ the bound eigenfunctions decay exponentially to zero.
Second, for $\lambda\le\lambda_0$ the eigenfunctions are unbound and correspond to 
``traveling particles''. The unbound eigenmodes have a continuous spectrum of their 
growth rates, $0<\lambda\le\lambda_0$. The unbound eigenfunctions 
asymptotically become a mixture of cosine and sine standing waves as $x\to\infty$. 
Eigenvalue $\lambda_0$ corresponds to the fastest growing unbound eigenmode.
The properties of the magnetic field amplified by kinematic dynamo action are fully 
determined by all growing eigenmodes of the field. In particular, the magnetic energy 
spectral function $F_B(k,t)$ is a linear sum of all energy spectral eigenfunctions, 
and the same is true for the electric current helicity spectral function $H_B(k,t)$.
Thus, to find the properties of the growing magnetic field, it is sufficient to find 
growth rates and spectral eigenfunctions of all field eigenmodes.\footnote{
While the energy spectrum $F_B(k,t)=\langle|{\bf B}({\bf k},t)|^2\rangle/2$ is positive, 
the energy spectral eigenfunctions may be negative.}

\section{Results and conclusions}

In this work we choose Gaussian velocity correlation tensor~(\ref{KAPPA}),
\beq
\kappa_L=e^{-x^2}, \quad
g=\frac{4h\sqrt{2e}}{27\sqrt{3}}\left(5-\frac{4x^2}{3}\right)e^{-2x^2/3},
\label{GAUSSIAN_V}
\eeq
where $h$ is the kinetic helicity parameter, for which the condition $-1\le h\le1$ must be 
satisfied.\footnote{
Functions $\kappa_L(x)$ and $g(x)$ can not be chosen arbitrarily, their Fourier images
$G(k)$ and $F(k)$ must satisfy the realizability condition $|G(k)|\le F(k)/k$ (Moffatt 1978),
which results in $-1\le h\le1$. Analogously, functions $M_L(x,t)$ and $K(x,t)$ are restricted by 
the condition $|H_B(k,t)|\le 2kF_B(k,t)$.
}
Without loss of generality, the choice~(\ref{GAUSSIAN_V}) means that all velocity fluctuations
are of order unity, $v_0\sim 1$, and are on the scale of order unity, $l_0\sim1$. Thus, the 
Reynolds number is of order unity and the turbulence has single-scale turbulent eddies.

\begin{figure}[t]
\vspace{5.7truecm}
\includegraphics{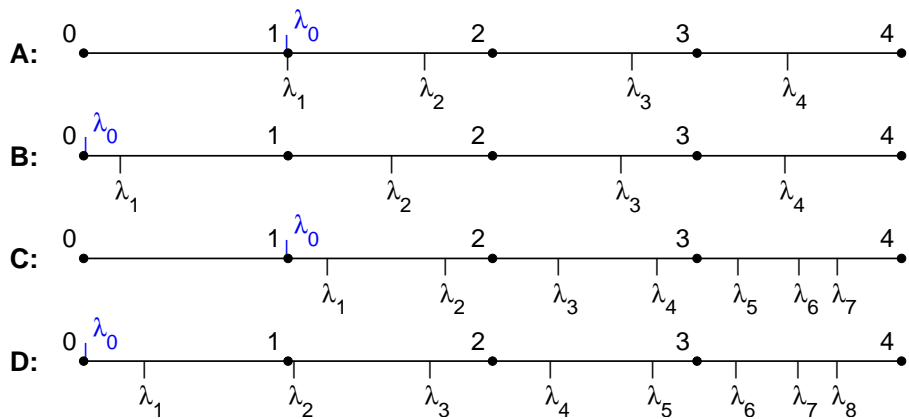}
\caption{Growth rates $\lambda_n$ of the bound magnetic eigenmodes, and $\lambda_0$ of
the fastest growing unbound eigenmode. Plots~A and~B are for $h=1$ and $0.01$ respectively, 
while $\eta=10^{-6}$ (${\rm Rm}\sim10^6$). Plots~C and~D are for $h=1$ and $0.01$ respectively, 
while $\eta=10^{-12}$ (${\rm Rm}\sim10^{12}$).
\label{FIGURE_1}
}
\end{figure}

\begin{figure}[t]
\vspace{6.0truecm}
\includegraphics{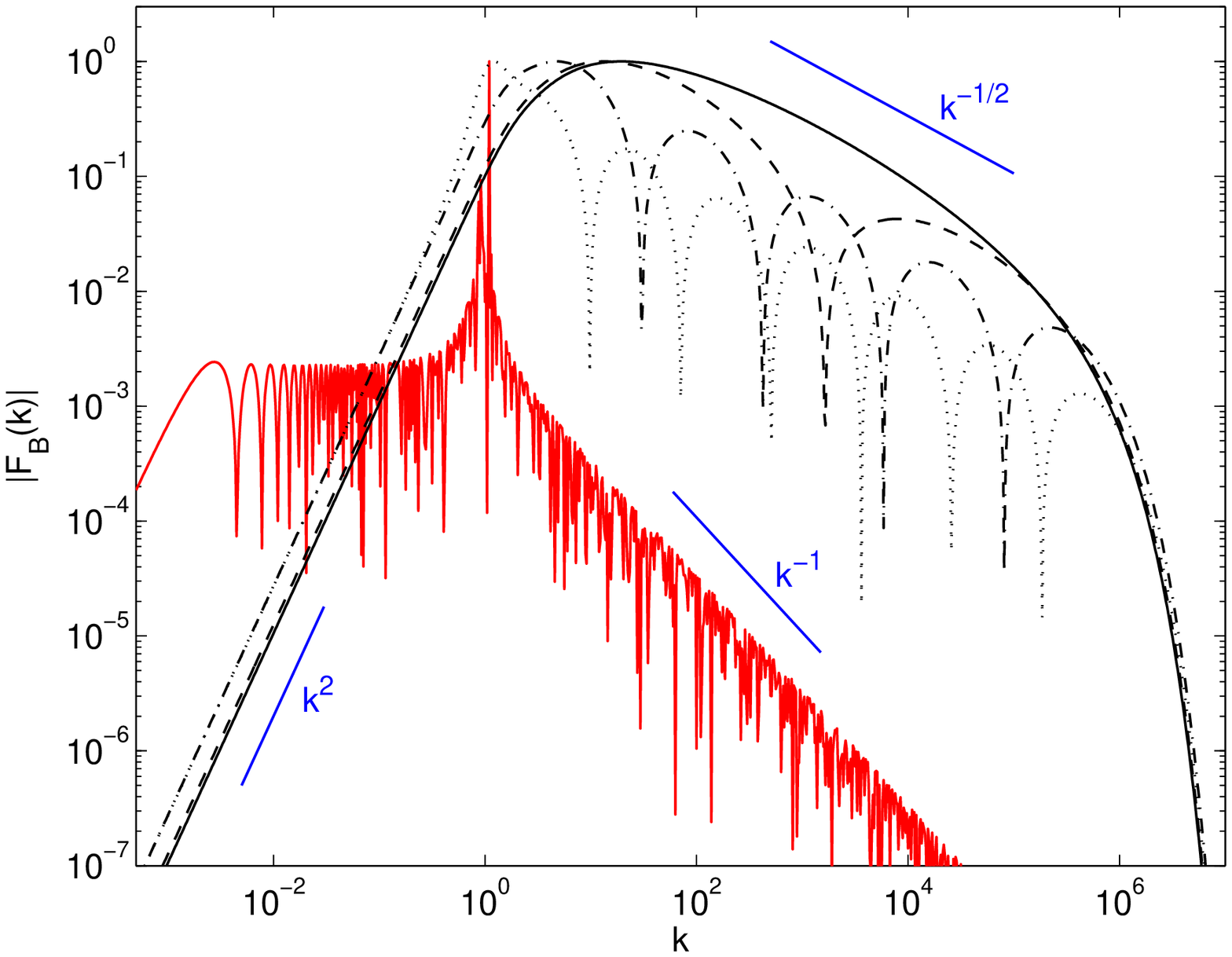}
\includegraphics{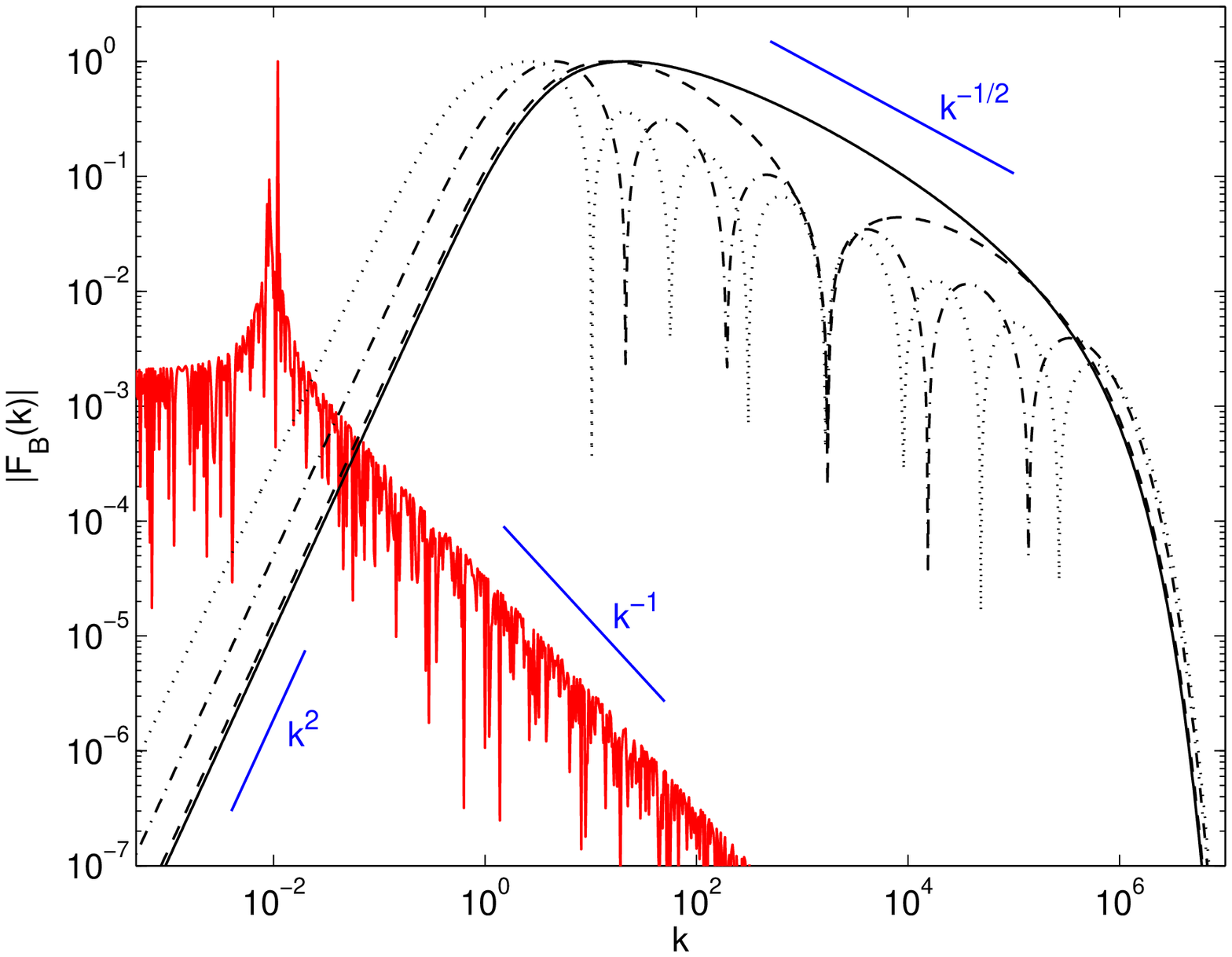}
\caption{Absolute values of magnetic energy spectral eigenfunctions for four selected bound
eigenmodes (shown by the dotted, dash-dot, dashed and smooth solid lines), 
and for the fastest growing unbound eigenmode (shown by the red jagged spiky solid 
lines). All plots are for $\eta=10^{-12}$ (${\rm Rm}\sim10^{12}$). 
The left and right plots are for the cases $h=1$ and $h=0.01$ respectively.
\label{FIGURE_2}
}
\end{figure}

\begin{figure}[t]
\vspace{6.0truecm}
\includegraphics{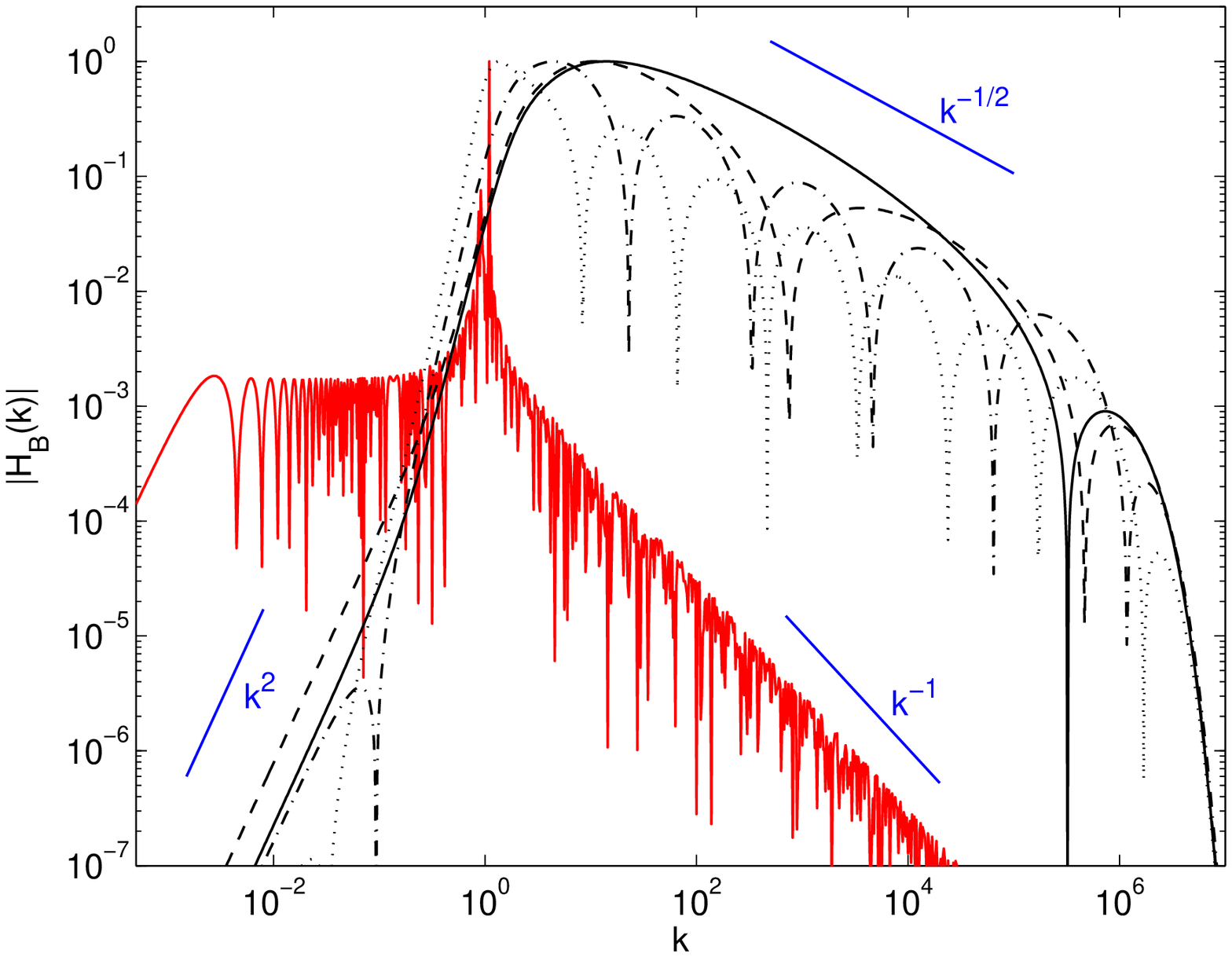}
\includegraphics{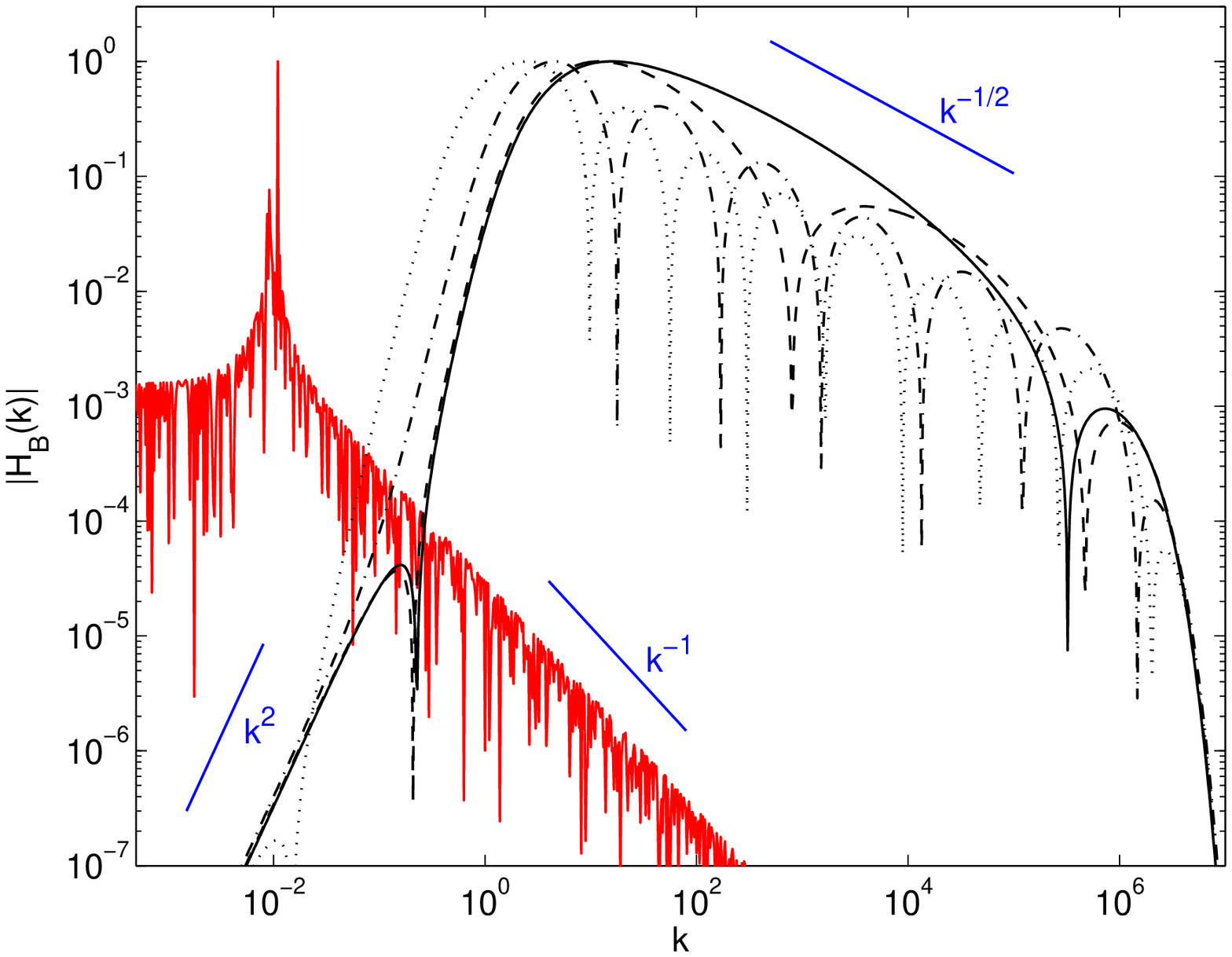}
\caption{The same as figure~\ref{FIGURE_2} except the absolute values of electric current 
helicity spectral eigenfunctions are plotted here.
\label{FIGURE_3}
}
\end{figure}

We study two cases for the magnetic Reynolds number,  ${\rm Rm}$. In the first case, 
${\rm Rm}\sim 10^6$, which is achieved by choosing magnetic diffusivity $\eta=10^{-6}$.
In the second case, ${\rm Rm}\sim 10^{12}$, which corresponds to our choice $\eta=10^{-12}$.
These two cases for the ${\rm Rm}$ value are considered in combination 
with two cases for the kinetic helicity: first, a case when $h=1$ in 
equation~(\ref{GAUSSIAN_V}) and the kinetic helicity is large, and, second, a case 
when $h=0.01$ and the kinetic helicity is small. Thus, in total we consider four cases for 
our choices of magnetic Reynolds number ${\rm Rm}$ and kinetic helicity parameter $h$.
The resulting growth rates $\lambda_n$ of the bound (localized) eigenmodes and $\lambda_0$ 
of the fastest growing unbound (non-localized) eigenmode are shown in figure~\ref{FIGURE_1}. 
The growth rates are measured in the units of a turbulent eddy turn-over rate $\sim v_0/l_0\sim 1$. 
For the cases ${\rm Rm}\sim 10^{12}$ and $h=1,\,0.01$, the 
logarithmic-scale plots of the absolute values of magnetic energy spectral eigenfunctions are 
given in figure~\ref{FIGURE_2}. The absolute values of the corresponding current helicity 
spectral eigenfunctions are plotted in figure~\ref{FIGURE_3}. 

In the case ${\rm Rm}\sim 10^6$ and $h=1$ there exist four bound magnetic field eigenmodes. They
have growth rates 
$\lambda_1\simeq0.9991$, $\lambda_2\simeq1.669$, $\lambda_3\simeq2.681$ and $\lambda_4\simeq3.443$, 
which are shown on the plot~A in figure~\ref{FIGURE_1}. The fastest growing unbound eigenmode 
grows at a rate $\lambda_0\simeq0.9943$. 
In the case ${\rm Rm}\sim 10^6$ and $h=0.01$ the number of bound eigenmodes is again four, 
$\lambda_1\simeq0.1750$, $\lambda_2\simeq1.507$, $\lambda_3\simeq2.629$ and $\lambda_4\simeq3.430$, 
while the fastest growing unbound eigenmode is $\lambda_0\simeq9.943\times10^{-5}$, refer to the 
plot~B in figure~\ref{FIGURE_1}.
Next, in the case ${\rm Rm}\sim 10^{12}$ and $h=1$ there are seven bound eigenmodes, 
all shown on the plot~C in figure~\ref{FIGURE_1}. Among these we select four bound modes
$\lambda_1\simeq1.192$, $\lambda_3\simeq2.322$, $\lambda_6\simeq3.499$, $\lambda_7\simeq3.686$, 
and we plot their spectral eigenfunctions by the dotted, dash-dot, dashed and smooth solid lines 
respectively on the left plots in figures~\ref{FIGURE_2} and~\ref{FIGURE_3}. The spectral 
eigenfunctions of the fastest unbound mode $\lambda_0\simeq0.9943$ are shown by the red jagged 
spiky lines on these left plots.
Finally, in the case ${\rm Rm}\sim 10^{12}$ and $h=0.01$ there are eight bound eigenmodes, 
refer to plot~D in figure~\ref{FIGURE_1}. The spectral eigenfunctions of four selected bound modes 
$\lambda_1\simeq0.2948$, $\lambda_3\simeq1.694$, $\lambda_7\simeq3.494$ and $\lambda_8\simeq3.685$ 
are shown by the dotted, dash-dot, dashed and smooth solid lines on the right plots in 
figures~\ref{FIGURE_2} and~\ref{FIGURE_3}. The spectra of the fastest unbound mode 
$\lambda_0\simeq9.943\times10^{-5}$ are again shown by the red jagged spiky lines.
As an interesting result, note that in figures~\ref{FIGURE_2} and~\ref{FIGURE_3} the envelopes 
of all bound eigenfunctions -- not only the envelope of the fastest growing eigenfunction -- 
agree with the Kazantsev spectral slope $k^{-1/2}$ (compare to Kulsrud and Anderson 1992).

Using the results presented in figures~{\ref{FIGURE_1}--\ref{FIGURE_3}}, we make the 
following conclusions. 
First, the value of kinetic helicity parameter $h$ does not have much effect on the fastest 
growing bound eigenmodes (e.g., compare modes $\lambda_5$, $\lambda_6$, $\lambda_7$ on plot~C 
to modes $\lambda_6$, $\lambda_7$, $\lambda_8$ on plot~D in figure~\ref{FIGURE_1}), but it 
does influence the unbound eigenmodes. This result is expected because helicity of velocity 
fluctuations does not propagate to small scales (Kulsrud and Anderson 1992), however, 
$\lambda_0=g^2(0)/[\kappa_L(0)+2\eta]\sim h^2v_0/l_0$.
Second, at a fixed kinetic helicity value, when the magnetic Reynolds number increases, the 
number of bound modes increases (refer to figure~\ref{FIGURE_1}). This is also expected 
because, the diffusion of magnetic field decreases with an increase of the magnetic Reynolds 
number.
Third, let us refer to the spectra of the fastest growing unbound eigenmodes $\lambda_0$, which
are shown by the the red jagged spiky solid lines in figures~\ref{FIGURE_2} and~\ref{FIGURE_3}.
We see that when the value of the kinetic helicity drops by a factor of a hundred (from $h=1$ to 
$h=0.01$), the location of the peaks of these spectra shifts to larger scales by the same factor. 
Thus, the characteristic scale of eigenmode $\lambda_0$ is approximately equal to $\sim l_0/h$, 
and this mode becomes large-scale when kinetic helicity is small, in agreement with the 
physical picture of large-scale dynamo action (Steenbeck \etal 1966).~\footnote{
At large correlation scales $x\to \infty$, the eigenfunction of eigenmode $\lambda_0$ 
asymptotically becomes a mixture of cosine and sine standing waves with wavenumber 
$k_0=\sqrt{\lambda_0}/\sqrt{\kappa_L(0)+2\eta}\sim h/l_0$.
}

We thank Fausto Cattaneo for useful discussions. This work was supported by the NSF Center 
for Magnetic Self-Organization in Laboratory and Astrophysical Plasmas. SB is supported by 
the U.S.~Dept.~of Energy under the grant No.~DE-FG02-07ER54932.

%------------------------------------------------------------------------------------------

\section*{References}
\begin{harvard}
\item[] Boldyrev, S., Cattaneo, F., and Rosner, R. 2005, Phys.~Rev.~Lett., {\bf 95}, 255001
\item[] Brandenburg, A., and Subramanian, K. 2005, Phys. Rep., 417, 1
\item[] Kazantsev, A.~P. 1968, JETP, {\bf 26}, 1031
\item[] Kim, E., and Hughes, D.~W. 1997, Phys. Lett.~A, {\bf 236}, 211
\item[] Kraichnan, R.~H., Phys. Fluids 1968, {\bf 11}, 945
\item[] Kulsrud, R.~M. 2005, {\em Plasma Physics for Astrophysics} (Princeton University Press)
\item[] Kulsrud, R.~M., and Anderson, S.~W. 1992, Astrophys. J., {\bf 396}, 606
\item[] Kulsrud, R.~M., and Zweibel, E. G. 2008, Rep. Prog. Phys., {\bf 71}, 046901
\item[] Lynden-Bell, D. (ed.) 1994, {\em Cosmical Magnetism} (Dordrecht: Kluwer Academic Publishers)
\item[] Malyshkin, L., and Boldyrev, S. 2007, Astrophys. J. Lett., {\bf 671}, L185
\item[] Moffatt, H.~K.~1978, {\em Magnetic Field Generation in Electrically Conducting Fluids} 
        (Cambridge U.~Press)
\item[] Monin, A.~S., and Yaglom, A.~M. 1971, {\em Statistical Fluid Mechanics} (MIT Press)
\item[] Parker, E. N. 1979, {\em Cosmical Magnetic Fields} (Oxford: Clarendon Press)
\item[] Steenbeck, M., Krause, F., and Radler, K.~H. 1966, Z. Naturforsch., {\bf 21}a, 369
\item[] Vainshtein, S.~I., and Kichatinov, L.~L. 1986, J. Fluid Mech., {\bf 168}, 73
\item[] Vainshtein, S. I., and Zeldovich, Ya. B. 1972, Sov.~Phys.~Uspekhi, {\bf 15}, 159
\item[] Zweibel, E.~G., and Heiles, C. 1997, Nature, {\bf 385}, 131
\end{harvard}

%------------------------------------------------------------------------------------------

\clearpage

\end{document}